\documentclass[lettersize,journal]{IEEEtran}
\usepackage{amsmath,amsfonts}
\usepackage{array}
\usepackage{textcomp}
\usepackage{stfloats}
\usepackage{url}
\usepackage{verbatim}
\usepackage{graphicx}
\usepackage{CJKutf8}

\def\BibTeX{{\rm B\kern-.05em{\sc i\kern-.025em b}\kern-.08em
    T\kern-.1667em\lower.7ex\hbox{E}\kern-.125emX}}
\usepackage{balance}

\usepackage{framed}

\usepackage{graphicx}
\usepackage{balance}  
\usepackage{epsfig}
\usepackage{graphicx}
\usepackage{epstopdf}
\usepackage{latexsym}
\PassOptionsToPackage{noend}{algpseudocode}
\usepackage{algpseudocode}
\usepackage{pifont}
\usepackage{epsfig}
\usepackage{amssymb}
\usepackage{amsmath}
\usepackage{amsfonts}
\usepackage{subfigure}
\usepackage{stmaryrd}
\usepackage{url}
\usepackage{multirow}
\usepackage{array}
\usepackage[normalem]{ulem}
\usepackage{color}
\usepackage{cleveref}
\usepackage{xspace}
\usepackage{mathtools}
\usepackage{soul}
\usepackage{listings}
\usepackage{enumitem}
\usepackage{xcolor}
\usepackage{tikz}
\usepackage{ragged2e}
\newsavebox{\blackball}
\newsavebox{\greenball}

\usepackage[utf8]{inputenc}
\usepackage{enumitem}
\usepackage{amsmath}

\usepackage{bbding}
\usepackage{amssymb}
\usepackage{epstopdf}
\usepackage{epsfig,endnotes}
\usepackage{url}
\usepackage{array}
\usepackage{booktabs}
\usepackage{threeparttable}
\usepackage{tabularx}
\usepackage{tabulary}

\usepackage{CJKutf8}
\usepackage[most,breakable]{tcolorbox}
\usepackage{etoolbox}
\usepackage{fancyhdr}
\usepackage{lipsum}
\usepackage[fencedCode]{markdown}

\usepackage{wasysym}

\usepackage[lined,boxed,vlined,ruled,linesnumbered]{algorithm2e}

\usepackage{bm}

\newcolumntype{M}[1]{>{\centering\arraybackslash}m{#1}}
\errorcontextlines\maxdimen
\newcommand{\hi}[1]{\vspace{.25em} \noindent {\bf #1}\xspace}

\newcommand{\leadingtext}[1]{\vspace{.25em}\noindent{\textbf{#1}}}

\lstdefinestyle{jsonStyle}{
	basicstyle=\small\ttfamily,
	columns=fullflexible,
	showstringspaces=false,
	commentstyle=\color{codegreen}\upshape,
	stringstyle=\color{codegreen},
	morestring=[b]",
	moredelim=[s][\color{codepurple}]{\{}{\}},
	moredelim=[s][\color{codepurple}]{[}{]},
	moredelim=[l][\color{codepurple}]{:},
	moredelim=[l][\color{codepurple}]{,}
}

\newcommand{\oursys}{\textsc{DataAgent}\xspace}

\newcommand{\cmark}{\CIRCLE}        
\newcommand{\warn}{\LEFTcircle}   
\newcommand{\xmark}{\Circle}       

\definecolor{codegreen}{rgb}{0,0.6,0}
\definecolor{codegray}{rgb}{0.5,0.5,0.5}
\definecolor{codepurple}{rgb}{0.58,0,0.82}
\definecolor{backcolour}{rgb}{0.95,0.95,0.92}

\newcommand{\zxh}[1]{\textcolor{red}{ #1}}

\tcbuselibrary{most} 
\newtcolorbox{Box4}[2][]{enhanced,
    arc=2mm,lower separated=false,
    colback=gray!20,colframe=white,
    fonttitle=\bfseries,
    colbacktitle=white!30!gray,
    coltitle=black,
    attach boxed title to top left={
        xshift=2cm,
        yshift=-2mm},
    title=#2,#1
}

\begin{document}
\begin{CJK*}{UTF8}{gbsn}

\linespread{0.94}

\pagestyle{plain}

\pagenumbering{roman}



\clearpage
	
\pagestyle{plain}

\title{Data Agents: Agentic Data Systems}

\author{Guoliang Li, Peiyao Zhou, Xuanhe Zhou, Ji Sun, Yuyu Luo, Ju Fan
\thanks{Guoliang Li, Peiyao Zhou, and Ji Sun are with the Department of Computer Science, Tsinghua University, Beijing, China. E-mail: liguoliang@tsinghua.edu.cn; Xuanhe Zhou is with Shanghai Jiao Tong University, Shanghai, China. {Yuyu Luo is with The Hong Kong University of Science and Technology (Guangzhou), China.} {Ju Fan is with Renmin University of China, Beijing, China.}}
\thanks{Corresponding author: Guoliang Li}\vspace{-2em}
}

\markboth{IEEE Transactions on Knowledge and Data Engineering, Vol. 0, No. 0, January 2025}{Li: Data Agents: Agentic Data Systems}

\pagestyle{plain}

\renewcommand\thesection{\arabic{section}}
\setcounter{section}{0}

\pagenumbering{arabic}
\setcounter{page}{1}
\setcounter{figure}{0}
\setcounter{table}{0}


\maketitle


\begin{abstract}
Traditional data systems face profound limitations in the AI era, relying on human-crafted pipelines, lacking semantic understanding of heterogeneous data, and operating through rigid, reactive processing. To address these challenges, we propose a new paradigm called the Data Agent, designed to manage, process, and analyze data with minimal human intervention. Data agents autonomously execute a wide range of data-related tasks, transforming traditional data systems by shifting from manual design to autonomous orchestration, from literal manipulation to semantic interpretation, and from reactive to proactive processing. Our Data Agent system includes six components: semantic data organization, semantic operators, agentic pipeline orchestration and optimization, feedback-driven refinement, memory management, and proactive adaptation. Building on this foundation, we also develop two specialized agents: the data analytics agent and the data science agent. Experiments on real benchmarks demonstrate significant performance gains of our data agent over state-of-the-art methods. We identify open challenges to guide future research in building fully autonomous data systems.
\end{abstract}


\maketitle


\section{Introduction}
\label{sec:intro}

Data systems that encompass the collection, storage, processing, management, and analysis of data have played vital roles in numerous data-related applications~\cite{bigdatasurvey,llmdata,li2025data,DBLP:journals/corr/abs-2510-23587}. However, traditional data systems, including conventional relational databases (e.g., PostgreSQL and MySQL), data warehouses, and manually orchestrated data pipelines built with tools (e.g., Airflow), face limitations in the AI era when supporting autonomous and intelligent applications~\cite{sun2026agenticdatabench,DBLP:journals/pvldb/LiLCLT24}. Although modern cloud providers offer services that automate data infrastructure scaling and execution (e.g., AWS Glue for ETL, SageMaker for ML, and Bedrock for LLM integration), these platforms still largely require humans to define the processing logic, configure pipelines, and specify data schemas. In other words, they automate execution but lack semantic data understanding and intent-driven autonomous orchestration.

\leadingtext{Limitation 1: Manually-Designed Pipelines.} Most current data systems depend on manually-designed pipelines that are tailored to specific tasks, such as building data science pipelines~\cite{datajuicer,dataanalysischange,deepanalyze,stage3ds} or facilitating interactive data analytics~\cite{xiong2024interactivekbqa,kunjir2020black,herodotou2011starfish}. These designs result in high adaptation costs when transitioning to new domains and use cases, significantly restricting the scalability and flexibility of the data system. For example, traditional data cleaning pipelines usually consist of several manual steps: identifying relevant data, applying transformation rules (such as normalizing numerical values and standardizing date formats), validating data consistency, and managing exceptions or missing values~\cite{refinedweb,fineweb, DBLP:conf/sigmod/Chai0FL23,DBLP:journals/pvldb/YangLCFCT25,DBLP:journals/pacmmod/LanWBLJLLHQH26}. Each step requires hand-coded rules and deep domain knowledge, which are intricately connected to the specific dataset or task. Consequently, when {\it migrating a pipeline designed for financial transaction data to healthcare patient records}, substantial manual modifications are necessary to account for differences in data schema, formats, and domain-specific rules. This high adaptation cost greatly limits the scalability and flexibility of such systems.

\leadingtext{Limitation 2: Limited Semantics Capabilities.} Existing data systems struggle to grasp the complexity and semantics of real-world heterogeneous data, which includes both structured and unstructured data. They operate under a {query-time completeness assumption} that {treats the stored dataset as the complete and authoritative source of truth.} 
Consequently, these systems fall short in managing diverse, noisy, or semantically rich data environments that require adaptive understanding and reasoning rather than fixed logic. For example, analyzing data from a data lake requires interpreting complex nested structures (e.g., multi-row/column headers and merged cells) and semantically dense entries such as product descriptions~\cite{tang2026straptor}. Similarly, financial reports demand reasoning over heterogeneous structures, inconsistent terminology, and context-dependent content~\cite{aop}. Current systems often cannot align semantically equivalent domain concepts or infer the intent behind aggregated records, making effective heterogeneous data processing highly challenging.


\leadingtext{Limitation 3: Reactive Processing.} Traditional data systems are limited to reacting to predefined queries or reports and lack the ability to proactively identify trends, anomalies, and necessary updates. This proactive capability is crucial for enabling organizations to anticipate changes and respond more strategically. For instance, in a reactive analytics scenario, a company might run reports at the end of each month to analyze sales data. If the analysis reveals that certain products have consistently underperformed, the company would respond by adjusting future orders or launching targeted promotions. However, by the time these insights are gathered and acted upon, the company may have already incurred losses from overstocking or missed opportunities to capitalize on market trends. Transitioning from reactive processing to proactive processing presents a significant challenge.

\begin{figure*}[!t]
	\vspace{-2em}
	\centering
    \includegraphics[width=\linewidth, trim={0 0 0 0},clip]{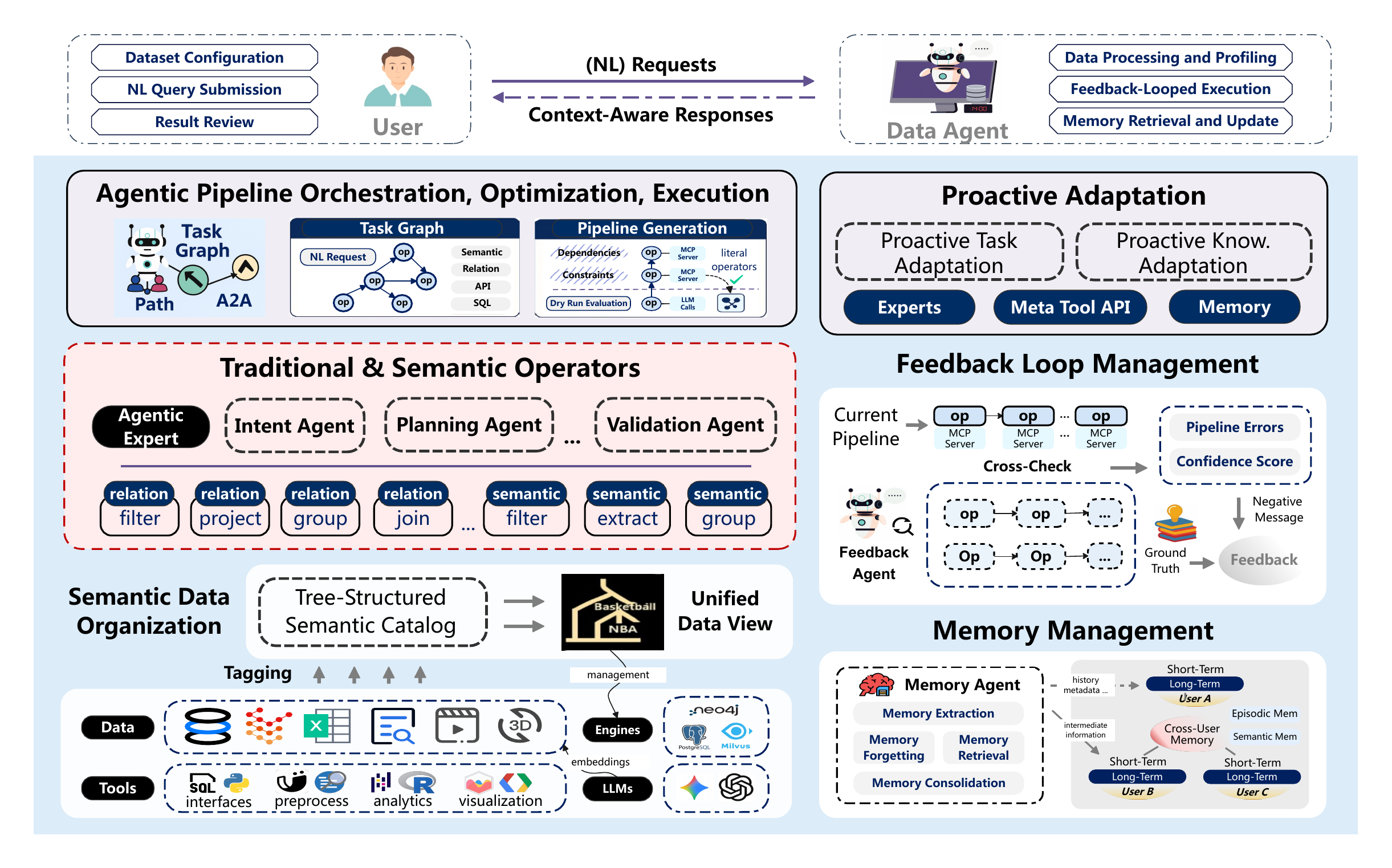}
	\vspace{-2.5em}
	\caption{Data Agent Architecture. }
	\label{fig:arch}
	\vspace{-1.5em}
\end{figure*}

\leadingtext{Paradigm Shift.} To overcome these limitations, \textbf{we propose a new paradigm called the \textsf{Data Agent}, which autonomously and proactively discovers, acquires, manages, processes, and analyzes data}. It executes data tasks to achieve a closed-loop system, moving from data to decision-making with minimal human intervention. Data agents transform traditional data systems by shifting from {\it manual design} to {\it autonomous orchestration}, from {\it literal manipulation} to {\it semantic interpretation}, and from {\it reactive processing} to {\it proactive processing}. Data agents have the potential to revolutionize a wide range of data-related tasks, from data preparation to analysis and decision-making. For example, when a company detects a revenue drop at a retail chain, traditional data systems simply report that ``sales have declined''. In contrast, a Data Agent proactively generates hypotheses, such as competitor promotions, severe weather, or declining delivery ratings. Then it automatically orchestrates a semantic pipeline to retrieve external evidence from sources such as weather APIs, review platforms, and competitors' websites. Afterward, it autonomously conducts deep semantic understanding, integrated analysis, and reasoning, ultimately generating decision-making recommendations for timely actions. Note that unlike general AI agents that still require significant human oversight and intervention for designing and coding data-related tasks~\cite{agentsurvey1,agentsurvey2,agentsurvey3,luo2025nvbench}, Data Agents operate with a high level of autonomy for performing end-to-end data tasks with minimal human intervention, creating a self-driven process from data collection to decision making.

\newcolumntype{C}{>{\Centering\arraybackslash}X}

\begin{table*}[!t]\vspace{-2em}
\centering
\caption{Comparison of Data Agents with Typical Automation Paradigms.}
\vspace{-1em}
\label{tab:positioning}
\small
\renewcommand{\arraystretch}{1.4} 
\begin{tabularx}{\textwidth}{@{} >{\RaggedRight\bfseries}p{2cm} C C C C C @{}}
\toprule
Dimension & Industrial Systems & LLM Orch. Stacks & Workflow Compilers & NL2SQL Systems & Data Agents \\
\midrule

Core \newline Functionality & 
ETL / Catalog / \newline Scheduling & 
Tool calling / \newline RAG / Agents & 
DAG \newline Compilation & 
NL-to-SQL & 
Autonomous semantic analytics pipeline orchestration \\
\midrule

Autonomous End-to-end Orchestration & 
\xmark \newline Human-defined \newline workflows & 
\xmark \newline Human-defined \newline logic & 
\xmark \newline Human-defined \newline DAG & 
\cmark & 
\cmark \newline NL-to-pipelines; \newline Feedback-driven \\
\midrule

Semantic-Aware \newline Processing & 
\xmark & 
\warn \newline Without data-centric \newline optimization & 
\xmark & 
\xmark \newline SQL operations \newline only & 
\cmark \newline Semantic data \newline organization \\
\midrule

Proactive \newline Adaptation & 
\xmark & 
\xmark & 
\xmark & 
\xmark & 
\cmark \newline Evolving memory \newline and data profiles \\
\midrule

Data-Centric \newline Memory & 
\xmark \newline Execution \newline state only & 
\warn \newline Generic \newline conversational & 
\xmark \newline DAG runtime \newline state only & 
\warn \newline Query-session \newline context only & 
\cmark \newline Working, Episodic, \newline Semantic memory \\
\bottomrule
\end{tabularx}\vspace{-1em}
\end{table*}

To better position the \textsf{Data Agent} paradigm beyond existing data automation systems and general-purpose agent frameworks, we compare it with industrial platforms (e.g., AWS Glue), generic LLM orchestration stacks (e.g. LangChain), workflow compilers (e.g. AutoML) and existing NL2SQL systems in Table~\ref{tab:positioning}. Specifically, we identify four categories of existing systems: (i) \emph{Industrial systems} (e.g., AWS Glue), which provide ETL, cataloging, and scheduling support but still require human-defined workflows; (ii) \emph{LLM orchestration stacks} (e.g., LangChain and AutoGen), which support tool calling and RAG but lack native semantic data organization, data-centric optimization, and specialized data operators; (iii) \emph{Workflow compilers} (e.g., AutoML, Apache Airflow), which compile human-defined DAGs but require manual pipeline specifications; and (iv) \emph{NL2SQL systems} (e.g., DIN-SQL~\cite{DBLP:conf/nips/PourrezaR23}, DAIL-SQL~\cite{DBLP:journals/pvldb/GaoWLSQDZ24}), which focus primarily on structured schema mapping via NL-to-SQL translation, lacking multi-step tool orchestration, feedback loops, and memory. The Data Agent paradigm is a fundamental architectural shift tailored specifically for data systems, integrating semantic understanding, autonomous end-to-end orchestration, data-centric memory, and proactive adaptation into a unified framework.


\leadingtext{Challenges.} There are numerous challenges involved in transitioning from {\it manual design} to {\it autonomous orchestration}, from {\it literal manipulation} to {\it semantic interpretation}, and from {\it reactive processing} to {\it proactive processing}.

\leadingtext{Challenge 1. Autonomous Pipeline Orchestration.} Directly using LLMs to orchestrate a pipeline composed of relational and semantic operators for tasks described in natural language is impractical, because LLMs can produce hallucinations and lack domain-specific knowledge and up-to-date information. There are several challenges in orchestrating high-quality pipelines while minimizing LLM costs and latency. First, understanding the task, environment, underlying data, and domain knowledge presents a significant challenge. Second, evaluating the quality of the pipeline and iteratively improving it is challenging. Third, the abundance of tools and operators makes it costly to generate pipelines from a large number of possible operator combinations.

\leadingtext{Challenge 2. Semantic Interpretation.} Using LLMs directly for processing complex tasks like semantic interpretation and understanding is expensive. Lowering the costs associated with employing LLMs for processing semantic operations is a significant challenge.

\leadingtext{Challenge 3. Proactive Processing.} 
Proactive systems must be adaptive, continually updating their processes and models based on new data or changing conditions. Continuously predicting, learning, and adjusting to ensure that the system remains relevant and effective over time is challenging.

\leadingtext{Our Contributions.} To address these challenges, we propose a data agent framework. First, the \emph{Semantic Data Organization} component manages and links structured, unstructured, and semi-structured data. It builds semantic catalogs and indexes to facilitate the semantic retrieval of heterogeneous data. Second, the \emph{Semantic and Generative Operator} component enriches traditional literal operators with and LLM-driven reasoning and generation to support semantic-aware processing. Third, the \emph{Agentic Pipeline Orchestration and Optimization} component adopts a multi-agent strategy to interpret a given task described in natural language, decompose the task, orchestrate a pipeline, and then optimize and execute the pipeline. Fourth, the \emph{Feedback-loop Management} component evaluates the pipeline's quality (e.g., correctness and confidence) and delivers ground-truth based feedback for performance refinement. Fifth, the \emph{Memory Management} component maintains short-term memory for infrequent and temporary information, such as task context and execution logs, and long-term memory for retaining frequent information, like historical data, metadata, and orchestration feedback. It also designs a cross-user memory sharing mechanism to enhance user experience like cold-start. Sixth, the \emph{Proactive Management} component continuously predicts, learns, and adjusts the system to ensure that it remains relevant and effective over time. 


The data agent provides the essential foundation for autonomous data processing. Specialized data agents, such as data analytics agents and data science agents, depend heavily on this foundational layer to function effectively. We explore how to design data science and data analytics agents by leveraging the capabilities of the data agent.

We make the following contributions.

\noindent (1) We propose a system called \oursys, exploring a new paradigm for data systems that can autonomously handle data-related tasks with minimal human intervention.

\noindent (2) We propose effective techniques for designing a data agent, including semantic data organization, semantic operators, agentic pipeline orchestration, feedback loop management, short-term and long-term memory management, and proactive adaptation.

\noindent (3) We explore effective techniques for designing data analytics agents and data science agents.

\noindent (4) We have conducted extensive experiments using real benchmarks, and the results showed that our data agent significantly outperformed state-of-the-art techniques.

\noindent (5) We outline open problems to guide future directions.

The remainder of the paper is organized as follows. Section~2 presents the overall \oursys framework. Section~3 and Section~4 elaborate two key components, namely memory management and feedback management. Section~5 introduces the data analytics agent and data science agent. Section~6 evaluates our system, and Section~7 discusses open problems.

\section{Data Agent Framework}
\label{sec:overview}

This section presents a data agent framework (see Figure~\ref{fig:arch}) intended to autonomously manage data-related tasks.  Given a dataset, a task expressed in natural language (e.g., ``report average spending of customers who spent over \$1,000 on food last month''), and a predefined library of operators, including relational operators (such as filter, project, group by, and join) and semantic operators (like semantic filter, semantic extraction, and semantic grouping), the data agent autonomously orchestrates an executable pipeline composed of these operators. It selects and sequences the appropriate operators to match the intent of the natural language query, optimizes the pipeline for efficiency and correctness, and then executes it to produce the final result.


\leadingtext{Semantic Data Organization.} It organizes heterogeneous data and provides \textbf{a unified semantic layer} to support semantic access across diverse data sources, including structured, unstructured, and semi-structured data. First, to effectively identify semantically relevant data, \oursys constructs a unified semantic catalog across heterogeneous data sources using a tree-structured hierarchy. Each node in the tree represents a semantic label, such as basketball or NBA, while the edges between parent and child nodes denote category-subcategory relationships. Each leaf node connects to all semantically related nodes within that category. Second, to unify diverse data, \oursys generates embeddings for heterogeneous datasets and links data points that are highly similar in embedding space. Third, to efficiently identify relevant data, \oursys builds vector indexes on these embeddings and maintains a semantic catalog that maps data to semantically related categories. The semantic catalog is materialized in the shared memory layer, enabling all agents to access a consistent, structured view of data semantics rather than relying on isolated representations.  By utilizing this approach, \oursys can efficiently identify relevant data for a given task using the semantic catalog and vector indexes.  Specifically, for a given query, it utilizes sparse embeddings of the keywords along with dense embeddings of the entire query to search for relevant semantic categories within the semantic catalog. Using these semantic categories, it then identifies the pertinent data based on semantic indexes. 


\leadingtext{Relational and Semantic Operators.} Traditional relational operators (e.g., filter, join, and group-by) are effective for structured data but lack the semantic capabilities needed for semantic matching, entity recognition, and intent-level operations, especially over unstructured and semi-structured data. To address this limitation, \oursys incorporates a set of semantic operators, such as semantic filtering, semantic grouping, semantic joining, and semantic extraction, and integrates them with conventional relational operators. These semantic operators can be implemented through LLM invocation, LLM-generated programs, or embedding-based matching, depending on the task requirement and execution cost.

For unstructured and semi-structured data, \oursys directly applies semantic operators to perform content understanding and transformation. For structured data, \oursys can invoke existing system APIs (e.g., Spark and Flink) or issue SQL queries to database engines (e.g., PostgreSQL, ClickHouse, and DuckDB). These connections are established through Model Context Protocol (MCP) interfaces, which allow \oursys to natively integrate existing data systems while also supporting user-defined drivers and custom operators. Thus, the atomic operators in \oursys include semantic operators, relational operators, system APIs, and SQL queries. Given an NL task, the goal of \oursys is to autonomously translate the user intent into an executable pipeline composed of these atomic operators.

To reduce latency and cost while maintaining accuracy, \oursys executes semantic operators using a cascading strategy. First, it performs approximate nearest neighbor (ANN) retrieval over vector representations to obtain an initial candidate set. Second, it applies threshold-based pruning with predefined upper and lower confidence bounds to discard clearly irrelevant candidates and accept clearly relevant ones. Third, for the remaining ambiguous candidates, it invokes a lightweight LLM to conduct coarse-grained semantic filtering or constraint checking. Finally, a stronger LLM performs rigorous verification on the minimized subset and produces structured outputs, such as filtered rows, semantic join mappings, or extracted entities. This progressive design confines expensive LLM calls to the most ambiguous cases, thereby achieving a practical trade-off between effectiveness and efficiency.

\leadingtext{Agentic Pipeline Orchestration, Optimization and Execution.}  We design a multi-agent strategy to automatically generate a pipeline using the aforementioned operators, optimize pipelines based on multi-goal optimization (e.g., improve the quality, reduce the cost and latency). In this approach, each agent specializes in a particular sub-task, such as task understanding, task decomposition, data profile extraction, query planning, reasoning, and pipeline validation. These agents are coordinated through Agent-to-Agent (A2A) protocols, where agents exchange structured messages (e.g., JSON objects encoding task graphs, execution states, and operator signatures). For example, during task understanding and decomposition, high-level goals are parsed and transformed into executable sub-flows. During pipeline optimization, an agent optimizes the pipeline for various objectives like latency, cost, interpretability, and accuracy. When it comes to executing pipeline actions, the agent assembles reusable semantic operators, along with necessary tools or data sources, instead of functioning as monolithic black boxes. To illustrate how \oursys generates a pipeline, consider the NL query ``Which product experienced the largest decline in sales, and why?'', \oursys first uses the Semantic Data Organization component to retrieve relevant tables, reviews, catalogs, and indexes built offline, while also obtaining useful historical knowledge from memory. The orchestration component then generates an executable pipeline consisting of three steps: (1) an SQL generation operator over the \texttt{sales} table to identify the product with the largest monthly sales decline; (2) a semantic filtering operator over the \texttt{reviews} table to retain negative reviews related to the identified product; and (3) a semantic extraction and grouping operator to extract reported issues from the filtered reviews and group them by cause with counts. The generated pipeline combines structured analysis over relational data with semantic reasoning over textual evidence. Its accuracy is evaluated against task-specific reference outputs (e.g., returned records, numerical values, rankings, aggregated statistics). 

For pipeline orchestration, \oursys first converts the user's NL request into a structured task graph where nodes represent sub-goals, matched to candidate atomic operators (semantic, relational, API, SQL), and edges capture the potential execution and data dependencies (e.g., involving the same data sources). Upon the task graph, a planning agent decides an operator sequence by evaluating logical coherence and resource constraints (e.g., memory usage, batch size compatibility, execution locality). For non-deterministic operators like LLM calls, a validation agent further refines the plan by conducting low-cost dry runs (on sampled data or in simulated environments) for multiple candidate execution sequences in parallel, and dynamically selects the one with the most promising accuracy and latency. Through this cooperative process, \oursys autonomously synthesizes an optimized, executable pipeline. Note that if a pipeline is incorrect, \oursys uses feedback mechanisms to identify and correct errors, thereby improving its quality. The activation of different components is primarily event-driven. The execution of each node in the task graph triggers the activation of specific semantic or relational operators. Furthermore, events such as a low confidence score or a pipeline execution error trigger the Feedback Loop Management component (Section~\ref{sec:feedback}), which in turn reactivates the planning or validation agents to refine the pipeline. 

For pipeline optimization, we design a cost-based approach. \oursys estimates the cardinality of semantic operators using the semantic catalog, and applies standard techniques to estimate the cardinality of relational operators. Based on these estimates, \oursys uses a dynamic programming method to select the pipeline with the lowest expected cost.

For pipeline execution, we adopt a cascading strategy for semantic operators. Given a semantic operator, \oursys uses embedding models and lightweight LLMs to filter out irrelevant data. For each model, we define an upper threshold (items with similarity above this threshold are accepted as true matches) and a lower threshold (items with similarity below this threshold are rejected). \oursys prunes items below the lower threshold and verifies items whose similarities fall between the two thresholds using larger LLMs.

\leadingtext{Feedback Loop Management.}  Crucially, this feedback loop component provides effective mechanisms to enhance pipeline quality. First, \oursys verifies pipeline correctness through cross-check validations, which involve generating multiple pipelines and merging them using a majority-voting validation approach. For instance, it runs these multiple pipelines (each pipeline returns a set of results) and collects the results from each pipeline. Next, it ranks the results based on the frequency (number of pipelines that produce each result). For each pipeline, it calculates a ``coverage score" which is the sum of the frequencies of the results this pipeline produces. At last, it selects the pipeline with the highest coverage score. Second, \oursys assigns confidence scores to each pipeline and offers feedback for those with low scores. For example, it provides negative feedback for pipelines with a low ``coverage score''. Third, \oursys triggers warning signals and initiates corrective sub-flows upon detecting issues such as pipeline errors or conflicts. Fourth, \oursys injects data with ground truth and provides negative feedback for pipelines that make errors on this ground truth. Fifth, \oursys meticulously generates a data sample, uses an LLM to compute responses for it, and then evaluates the quality of the pipeline based on the LLM-generated responses. To achieve this, LLMs are used to generate data that satisfies the question and then execute the pipeline on the generated data. To prevent infinite loops, the system implements internal mechanisms including confidence thresholds and a maximum retry limit. The user acts as an optional, high-level validator: the system invokes user confirmation only when its internal confidence remains below a critical threshold after the maximum number of retries.

\leadingtext{Short-Term and Long-Term Memory Management.} Memory Management (MM) is strictly an internal component of the overall \oursys architecture. The user interacts with the \oursys interface, and the \oursys internally queries and updates the MM without requiring the user to act as an intermediary. To ensure continuity and adaptability, the \oursys uses short-term and long-term memory to store semantic knowledge, contextual metadata, and prior decisions. This enables it to reuse modeling templates, adapt to distribution shifts, and produce traceable, versioned artifacts. Short-term memory holds transient information -- such as task context, execution logs, ratings and feedback, and temporal patterns -- so only immediately relevant data is kept, with irrelevant or outdated items quickly discarded. Long-term memory stores persistent, frequently used information like historical data and metadata. To manage long-term memory efficiently, \oursys uses a hierarchical structure that layers memory by access frequency and relevance, and builds vector indexes to improve retrieval. It also employs efficient promotion mechanisms that extract useful items from short-term memory and persist them in long-term memory. Finally, it applies conflict-detection techniques to identify and resolve inconsistencies across memories, improving reliability.

\leadingtext{Proactive Adaptation.} Our system supports proactive adaptation to handle evolving tasks, shifting user intents, and concept drift in dynamic environments. Rather than reacting only after performance degrades, the system continuously monitors data agents using a set of predefined performance metrics, including query success rate, pipeline execution latency, result plausibility scores, user feedback signals. When significant deviations are detected (e.g., a sudden drop in accuracy or an increase in ambiguous queries), the system triggers an automated root-cause analysis: specialized diagnostic agents examine recent interactions, data distributions, and agent behavior to hypothesize potential causes—such as outdated knowledge, schema changes, or emerging user goals. These hypotheses are then verified through targeted validation (e.g., testing alternative pipelines on recent data or checking consistency with updated domain facts). Once confirmed, the system proactively fixes them, which may include updating memory priors, adjusting operator preferences, or replanning the pipeline -- often before the user even notices a problem. This closed-loop, anticipatory approach ensures the system remains robust, relevant, and responsive in the face of continuous change.

\leadingtext{Workflow.} \oursys shifts data systems from static pipelines to goal-driven, autonomous, adaptive processes. (1) The \emph{Semantic Data Organization} component discovers and links relevant structured, unstructured, and semi-structured data; (2) the \emph{Semantic Data Operators} component enriches data processing operations with LLM-based reasoning and generation; (3) the \emph{Agentic Pipeline Orchestration, Optimization and Execution} component interprets user intent, decomposes the task, composes a tailored execution plan, optimizes and executes the pipeline; (4) the \emph{Feedback Loop Management} component monitors pipeline correctness, verifies results, provides feedback, and adjusts future behavior; (5) the \emph{Memory Management} component maintains both short-term and long-term memory to store context and historical information;  and (6) the \emph{Proactive Adaptation} component anticipates new requirements by continuously learning from interactions. Specialized data agents, such as the data science agent and the data analytics agent, are built on top of the general data agent to provide capabilities tailored to specific data processing. These components are activated by specific events and perform corresponding actions (see Table~\ref{tab:event_component_activation}).  To support this process, we explicitly define two primary communication mechanisms among components. (1) \emph{Agent-to-Agent (A2A) protocols:} Agents (e.g., planning agent, execution agent, validation agent) in our system exchange JSON-formatted messages with predefined fields (e.g., \texttt{task\_id}, \texttt{input}, \texttt{output}, \texttt{status}). These messages encode task graphs, execution states, and operator signatures, enabling coordinated multi-agent workflows. (2) \emph{Shared Memory and Semantic Catalog:} The shared memory layer acts as a common data pool, where agents publish their intermediate results (e.g., extracted entities, filtered subsets) also in structured formats (e.g., DataFrames) rather than plain text, allowing downstream operators to process them programmatically. 



\begin{table*}[!t]\vspace{-2em}
\centering
\caption{Event-driven Activation Rules of System Components}
\label{tab:event_component_activation}\vspace{-1em}
\begin{tabular}{p{4.6cm}p{8.4cm}p{3.8cm}}
\toprule
\textbf{Event} & \textbf{Triggered Action} & \textbf{Activated Component} \\
\midrule
User query received 
& Retrieve relevant historical memory and prior task context 
& Memory Management \\

Query context initialized 
& Perform semantic data profiling over available sources 
& Semantic Data Organization \\

Query, memory, and profiles ready 
& Generate, optimize, and execute the analytical pipeline 
& Agentic Pipeline Orchestration \\

Pipeline execution completed 
& Verify execution correctness and detect runtime or semantic errors 
& Feedback Loop Management \\

Execution feedback available 
& Store useful intermediate plans and observations into short-term memory 
& Memory Management \\

Data missing error detected 
& Refine semantic profiles and update source understanding 
& Semantic Data Organization \\

Updated profiles available 
& Refine and regenerate the pipeline with improved grounding 
& Agentic Pipeline Orchestration \\

Pipeline runtime or logic error detected 
& Repair the pipeline structure and retry execution 
& Agentic Pipeline Orchestration \\

Short-term memory updated 
& Consolidate stable insights into long-term memory 
& Memory Management \\

Data sources updated 
& Refresh semantic catalogs, statistics, and retrieval indexes 
& Semantic Data Organization \\

Task completed 
& Anticipate likely follow-up analytical requirements 
& Proactive Adaptation \\
\bottomrule
\end{tabular}\vspace{-1em}
\end{table*}

\section{Memory Management in Data Agents}
\label{sec:memory}

The memory management component serves as the data agent's persistent knowledge repository, storing relevant data, past interactions, and learned insights accumulated throughout the workflow. It not only retains contextual information for continuity across tasks but also retrieves and supplies pertinent memories to inform and refine pipeline generation, enabling the data agent to adapt, improve, and make more informed decisions over time. 

\subsection{Data Agent Memory Categories}

The data agent achieves a balance between immediate responsiveness (via working memory) and persistent, scalable intelligence (via long-term memory). This architecture not only enhances coherence and personalization but also ensures factual accuracy and task efficiency over extended interactions.


\leadingtext{\bf Short-term Memory.} The working memory (usually short-term memory) serves as the model's contextual workspace, temporarily storing current conversation history, user interactions, and intermediate reasoning processes (such as chain-of-thought). It acts as the ``mental scratchpad'' or ``cognitive workbench'' for immediate reasoning tasks. However, it has limited capacity (constrained by context window length), volatile nature (cleared at session termination), dynamic and temporary storage. For example, in continuous dialogues, the LLM's ability to reference previously mentioned information relies on working memory.

\leadingtext{\bf Long-term Memory.}  To overcome the limitations of working memory, the agent leverages long-term memory, which resides externally (e.g., in vector databases or structured knowledge stores) and is selectively retrieved into working memory when needed. Long-term memory is categorized into three key types.

\leadingtext{\it (i) Episodic memory} enables personalized, context-aware interactions by capturing user-specific experiences, preferences, and events tied to particular individuals or sessions. It records what happened from the user's perspective (such as choices, feedback, or intentions) and uses this history to tailor future responses. For example, if a user says, ``I prefer PostgreSQL over other databases'', this preference is stored as an episodic memory and  used to guide database-related recommendations, ensuring continuity and personalization across interactions.

\leadingtext{\it (ii) Semantic memory} stores general, objective knowledge about the world—facts, concepts, domain rules, and commonsense reasoning that are not tied to any individual user. Its role is to ground the agent’s responses in accurate, verifiable information. For instance, when asked about ``PostgreSQL’s query rewrite rules'', the data agent retrieves authoritative documentation from an external knowledge base via semantic memory, ensuring factual correctness rather than relying solely on internal model parameters.

\leadingtext{\it (iii) Procedural memory} encodes how to do things: it preserves reusable skills, workflows, and step-by-step procedures for completing specific tasks. This memory type allows the agent to automate complex operations consistently and efficiently. For example, upon receiving a request for data analysis, the agent can invoke a stored procedural memory to automatically generate a standardized analytical report template, including data cleaning, visualization, and statistical summary steps, based on best practices or previously validated pipelines.

These long-term memories equip the data agent with the ability to remember personal context, reason with factual accuracy, and execute tasks reliably. In our implementation, each long-term memory entry is stored as a structured tuple with metadata (e.g., type, timestamp, importance score, embeddings), and conflict resolution is handled through a graph-based memory algorithm (Algorithm~\ref{alg:memory_insertion}) with specific resolution rules (recency-based, confidence-based, confirmation-based) as detailed in Section~\ref{subsec:mm:ops}. The version graph maintains full traceability of memory states, enabling the system to reconstruct the chronological evolution of any stored preference.






\leadingtext{\bf Transferring from Working Memory to Long-term Memory.} 
To transfer information from working memory to long-term memory, data agents follow a structured workflow. First, they identify what is worth remembering—filtering high-value, user-relevant, or task-critical content from the transient context (called memory extraction). Next, this information is encoded and structured into a standardized format, often enriched with metadata (e.g., memory type, timestamp, importance score) and converted into embeddings for semantic retrieval. The structured memory is then inserted in a persistent backend (memory insertion), such as a vector database or knowledge graph, ensuring durability across sessions. In practice, memory tuples stored in long-term memory may conflict with one another—either due to evolving user preferences, corrected facts, or contradictory experiences. To maintain consistency, the system employs memory update operators that detect and resolve such conflicts. 
Additionally, memories can become outdated or irrelevant over time, necessitating controlled forgetting. 
The system can automatically delete or archive such entries based on triggers like API deprecation notices, lack of recent use, or explicit user feedback. This memory deletion mechanism prevents the agent from relying on obsolete knowledge, keeping its behavior accurate and up to date. Finally, the system enables future retrieval (memory retrieval) by querying this long-term store when relevant -- pulling pertinent memories back into working memory to inform reasoning, personalize responses, or guide actions in new interactions.

\subsection{Data Agent Memory Operators} \label{subsec:mm:ops}

The data agentic memory system encompasses a full lifecycle of memory management, including:

(1) Memory extraction: identifying and distilling salient information from interactions or observations;

(2) Memory insertion, update, and deletion (i.e., forgetting): persisting new memories, revising conflicting or outdated ones, and removing obsolete entries;

(3) Memory retrieval: retrieving relevant memories based on semantic similarity, user context, or descriptions;

(4) Memory consolidation: periodically summarizing, compressing, or reorganizing memories to improve coherence, reduce redundancy, and enhance utility.



These capabilities enable the data agent to learn continuously, maintain consistent knowledge, and adapt over time.

\subsubsection{Memory Extraction}

The memory extraction phase identifies, distills, and evaluates valuable information from raw interaction logs or working memory to determine what should be preserved in long-term memory. This process is carried out through three specialized operators.

The summarization and condensation operator compresses lengthy dialogues or documents into concise, high-fidelity summaries that retain essential facts, decisions, and user intentions—dramatically reducing storage overhead while preserving meaning.  
The structured extraction operator precisely extracts entities (people, projects), attributes (preferences, specifications), events (decisions, changes) from unstructured text and structures them. Given the working memory content, \oursys invokes LLM function calling capabilities or pre-trained named entity recognition (NER) models, which output key-value pairs or knowledge triples for enriching episodic metadata or building semantic knowledge graphs. 


The value assessment operator predicts long-term potential value of information before storage, dynamically determining storage priority and duration. Given working memory content,  \oursys uses rule-based (high score for ``my preference'' mentions) combined with learned metrics (information entropy, relevance to core objectives, access frequency prediction), which outputs quantitative importance score for guiding storage, retrieval, and forgetting strategies.  {For example, during the execution of queries with a particular pattern, if the semantic filter is preferred more frequently and consistently achieves a higher positive feedback rate than the traditional filter, the system assigns a higher importance score to semantic filters based on both the recurrence and effectiveness of their filtered results. This memory is therefore retained for a longer period and prioritized during retrieval for similar queries that are frequent and effective. In contrast, the traditional filter  receives a lower score and is discarded or set to expire quickly.}

\subsubsection{Memory Insertion/Update/Deletion}

The memory management system uses insertion, update, and deletion (also called ``forgetting") operators to maintain the agent's memory as accurate, relevant, and efficient. Notably, the system often preserves version history, allowing past states to be audited or restored if needed. First, {\it Insertion} adds new memories to long-term storage. Second, {\it Update} manages changes and resolves conflicts between new and existing information. When conflicts arise (e.g., multiple values for the same key), the system uses rule-based strategies: $(i)$
Recency-based selection (``prefer the latest") for time-sensitive data, $(ii)$ Confidence-based selection using LLM-derived or retrieval scores, and $(iii)$
Explicit confirmation by invoking an LLM when ambiguity persists.
Conflicting entries are not overwritten but retained as parallel versions linked via a version graph, ensuring traceability. Conflicts are dynamically resolved during retrieval by selecting the most relevant version based on the query context, ensuring consistency and auditability. Third, {\it Deletion} removes or archives outdated, irrelevant, or low-value memories, similar to how humans naturally forget unimportant details. This prevents memory bloat and maintains system performance. Deletion follows three strategies: $(i)$ Value-driven: Periodically purges memories with importance scores below a threshold, $(ii)$
Time-driven: Removes memories after a set duration unless marked as high-value, and $(iii)$
Instruction-driven: Deletes memories in response to explicit user requests. 
Together, these operators ensure the agent’s memory remains accurate, adaptive, and efficient—retaining critical information while discarding the unnecessary.

\begin{algorithm}[!t]
\caption{Memory Insertion\&Conflict Resolution}
\label{alg:memory_insertion}
\KwIn{Extracted new memory $M_{new}$, Vector Index $\mathcal{I}$, Memory Version Graph $\mathcal{G} = (V, E)$}
\KwOut{Updated Memory Graph $\mathcal{G}$ and Index $\mathcal{I}$}

$v_{new} \gets \text{ComputeEmbedding}(M_{new})$\;
\tcp{Step 1: Locate relevant nodes}
$Anchors \gets \text{VectorSearch}(\mathcal{I}, v_{new}, \text{{top-$k$}})$\;
$ConflictNodes \gets \emptyset $\;
\vspace{0.1cm}
\tcp{Step 2: Detect Conflicts}
\ForEach{$node \in Anchors$}{
    \tcp{Find related memories or leaves}
    $RelatedNodes \gets \text{TraverseGraph}(\mathcal{G}, node)$\;
    \ForEach{$r \in RelatedNodes$}{
        $is\_conflict \gets \text{LLMJudgeConflict}(M_{new}, r)$\;
        \If{$is\_conflict$}{
            $ConflictNodes \gets ConflictNodes \cup \{r\}$\;
        }
    }
}
\tcp{Step 3: Resolve conflict}
$\text{AddNode}(\mathcal{G}, M_{new})$\;
\ForEach{$target \in ConflictNodes$}{
    $action \gets \text{ApplyRule}(M_{new}, target)$\; 
    \If{$action == \text{``Overwrite"}$}{
        $\text{Deactivate}(target)$\;
    }
    $\text{AddEdge}(\mathcal{G}, target \to M_{new})$\; 
}
$\text{UpdateIndex}(\mathcal{I}, M_{new})$\;
\Return{$\mathcal{G}, \mathcal{I}$}\;
\end{algorithm}



\subsubsection{Memory Retrieval/Search}

The memory retrieval operator is responsible for finding the most relevant memory segments from long-term storage to support the current task. To maximize recall and precision, \oursys uses a multi-stage retrieval pipeline that combines semantic, textual, and metadata-based signals. $(i)$ Semantic Retrieval: It computes vector embeddings for the current query and retrieves the {top-$k$} memory entries with the highest semantic similarity (e.g., using cosine similarity in a vector database). This captures meaning even when wording differs. 
$(ii)$ Textual Retrieval:  It performs keyword- or lexical-based matching (e.g., BM25) to find entries with the strongest textual overlap with the query. This helps catch precise matches that embedding models might smooth over. 
$(iii)$  Metadata Filtering: It narrows the candidate pool using structured metadata -- such as time range (e.g., only memories from the last 10 minutes), memory type (e.g., only procedural or user-specific episodic memories), or importance score (e.g., exclude low-priority entries).

After finding relevant memory segments, \oursys leverages the memory version graph (constructed during insertion) to search for relevant ones. Starting from the anchor nodes, it traverses adjacent and preceding nodes along the directed edges, and returns an ordered list of entries with timestamps. This allows the data agent to reconstruct the chronological evolution of a state or preference, resolving historical conflicts by tracing the active version paths.

Finally, \oursys fuses and reranks all candidates using a weighted scoring function that balances: semantic similarity, textual match strength, temporal recency, 
importance score (user preferences or verified facts rank higher). The result is a refined list of the {top-$k$} most relevant memory segments, ordered by overall relevance to the current context.


\subsubsection{Memory Consolidation}

The memory consolidation operator periodically refines stored memories, transforming fragmented experiences into richer, higher-level knowledge. It achieves this by compressing, abstracting, and linking raw memory entries, turning isolated events into meaningful patterns or generalized insights. For example, it may recognize recurring themes and create a meta-memory like: ``User is an advanced Python learner focused on performance optimization," which can guide more proactive and context-aware recommendations. Consolidation builds connections between previously unrelated memories, weaving them into a coherent knowledge network. By transforming episodic fragments into structured, interconnected knowledge, consolidation enables the data agent to reason more deeply, anticipate user needs, and deliver increasingly intelligent responses over time.

\subsection{Data Agent Memory Storage}

To enable efficient and semantically aware access to both recent context and historical knowledge, \oursys uses a unified memory storage architecture with distinct handling for working and long-term memory. Working memory that captures the immediate conversation history and task context is typically kept in a lightweight, short-lived buffer (e.g., in-session cache or token-limited prompt context). It is not persisted but periodically distilled into long-term memory. Long-term memory is persistently stored in a vector database. Each memory entry is encoded into a dense vector and enriched with structured metadata, including: Memory type (e.g., episodic, semantic, or procedural), Source (e.g., user statement, external knowledge base), Timestamp, and Importance score (based on relevance, user intent, or usage patterns). This design allows the agent to perform semantic search across large-scale memory collections. During inference, it retrieves the most relevant  information -- whether from recent interactions (via working memory) or accumulated experience (via long-term memory) -- to ground its reasoning, personalize responses, and maintain coherent, context-aware behavior over time.




\section{Feedback Management}
\label{sec:feedback}

The critical challenge of pipeline orchestration lies in verifying the correctness of the generated pipeline. Since ground-truth pipelines are often unavailable and errors can be subtle, we introduce an error detection and feedback mechanism to iteratively improve pipeline quality.  Our approach identifies and addresses two key types of pipeline errors.

(1) Syntactic errors: Invalid operator usage, incorrect parameter formats, or defective pipeline structures.

(2) Semantic errors: Logically incorrect transformations that produce results inconsistent with the user's intent (e.g., filtering on the wrong attribute).


\leadingtext{\bf Syntactic Error Feedback.} To detect syntactic errors in automatically generated data pipelines (e.g., invalid operator usage, incorrect parameter formats, defective structure), we employ a multi-layered validation strategy. First, schema-aware validation ensures all referenced columns exist and have compatible types. Second, operator signature checking verifies that each operator is called with the correct arguments, types, and required parameters. Third, pipeline grammar validation parses the pipeline as a structured program to confirm that it follows valid sequencing and compositional rules (e.g., no cycles, proper operator order). Additionally, a static dry run using metadata or mock data can catch  mismatches and runtime-like issues without running the full execution. Optionally, an LLM-guided self-check can provide heuristic feedback on ambiguous cases. In this way, these methods enable robust, automated detection of syntactic flaws before pipeline execution, improving reliability and reducing downstream failures.

\leadingtext{\bf Semantic Error Feedback.} We employ three  verification strategies to detect and correct semantic errors in pipelines.

\leadingtext{\it (i) Cross-Validation via Multiple Pipeline Generation.} To enhance robustness and reduce the risk of generating a single flawed interpretation, \oursys generates multiple diverse candidate pipelines for the same natural language (NL) query -- each representing a plausible way to translate the user's intent into executable operations. These candidates are then executed on the actual dataset, and their outputs are systematically compared using both quantitative and qualitative criteria (e.g., result cardinality, data distribution, logical consistency, and alignment with expected semantics). Pipelines that produce consistent, coherent, and semantically plausible results, especially those that converge with other high-quality candidates, are assigned higher confidence scores and prioritized. In contrast, pipelines yielding outlier results (e.g., empty outputs, implausible aggregates, or values inconsistent with the query's scope) are flagged as potentially erroneous and are either discarded or sent back for refinement through feedback loops. This cross-validation approach acts as a built-in consistency check, mimicking how multiple experts might independently solve the same problem and converge on a reliable answer.

\leadingtext{\it (ii) Synthetic Data Consistency Checking.} To proactively catch semantic or logical errors,  \oursys generates a small, controlled set of synthetic data that embodies key characteristics of the real dataset (such as schema structure, value ranges, and edge cases) but is simple enough to reason about manually or by LLMs. It then executes two parallel processes on this synthetic data: (1) the actual generated pipeline, and (2) a reference interpretation of the user's intent, derived either from the LLM’s step-by-step reasoning trace or a human-aligned expected behavior. By comparing the outputs of both, it assesses whether the pipeline faithfully implements the intended logic. If the results match closely -- in both content and structure --  it gains confidence in the pipeline’s correctness. However, if there is a mismatch (e.g., the pipeline returns all users while the expected result filters only high-value ones), it signals a semantic or logical flaw, such as misinterpreting a filter condition, applying an operation in the wrong order, or conflating similar concepts. This technique acts as a lightweight, interpretable ``unit test'' for pipeline validation, enabling early error detection without requiring full-scale execution or ground-truth labels.

\leadingtext{\it (iii) Pipeline-to-NL Consistency Evaluation.} To verify that the generated pipeline truly reflects the user's original intent, we employ a Pipeline-to-Natural-Language (Pipeline2NL) consistency check. Specifically, we automatically translate the executed pipeline back into a clear, human-readable natural language description. For example: ``Filter customers who spent over \$1,000 on food last month, then compute their average spending''. This reconstructed statement is then compared to the original user query (e.g., ``What's the average spending of customers who spent more than \$1,000 on food last month?'') using semantic similarity metrics such as embedding-based cosine similarity or LLM-powered entailment scoring.

A high similarity score indicates strong alignment between what the user asked for and what the pipeline actually does. Conversely, a low similarity score raises a red flag: it suggests a potential misinterpretation or implementation drift, such as filtering on the wrong category (e.g., ``total spending'' instead of ``food spending''), omitting a time constraint (``last month''), or performing an incorrect aggregation. By closing the loop from code back to language, this technique provides an interpretable, intent-aware validation layer that complements execution-based checks and helps catch subtle semantic errors that might otherwise go unnoticed.




\section{Data Agent Variants}
\label{sec:variants}


The data agent establishes the groundwork for autonomous data processing. Specialized data agents, like the data analytics agent and data science agent, depend on this foundation to deliver these essential capabilities. Beyond this, such data agents should also specify task requirements, design feedback mechanisms, and provide proactive specifications. 

\vspace{-.5em}

\subsection{Data Analytics Agent}
\label{sec:unstructured-data}

It supports data analytics on heterogeneous data, including both structured and unstructured data. Unlike structured queries (where schema, types, and relationships are predefined), unstructured data is often free-form and diverse in format. The challenge lies not only in the absence of structure but also in the ambiguity, heterogeneity, and lack of clear semantic anchors. Traditional data pipelines struggle to handle such irregular inputs at scale~\cite{bigdatasurvey,llmdata}. This represents a fundamental shift: from {\it retrieval-based document systems handling unstructured data and syntactic analytics on structured data} to {\it semantic analytics across heterogeneous data}. The data analytics agent not only locates relevant content, but also interprets, reasons, and synthesizes insights, advancing data analytics toward an agent-driven paradigm.

\leadingtext{Basic Idea.} The data analytics agent must access all data sources, build a semantic catalog for heterogeneous data, define data analytics-related semantic operators, and integrate with the APIs, tools, and SQL interfaces of the underlying systems. The data analytics agent then creates a pipeline for the specified task, optimizes it, and executes it.




\leadingtext{Agentic Pipeline Orchestration.} The data analytics agent starts by parsing the user query into a high-level semantic plan. It identifies the query intent (e.g., root cause analysis of a decline in economic indicators), selects relevant data (e.g., knowledge of possible causes and relevant data sources), chooses pertinent semantic operators (e.g., semantic filtering of relevant data and semantic grouping for the causes), and orchestrates a semantic pipeline. The pipeline orchestration proceeds iteratively, with LLMs assisting in interpretation, where data retriever  extracts relevant context from surrounding documents, and pipeline validator ensures correctness and consistency. When misalignment is detected, such as a pipeline error or low confidence level, the system reroutes the pipeline and updates its confidence scores, emulating human troubleshooting behavior. Moreover, to support such dynamic capability, the data analytics agent leverages its unified operator library, planning engine, and feedback-driven execution layer. These components interact with vector indexes, LLM APIs, and memory modules, enabling continual refinement and learning from past executions.  We also store crucial context information and historical data in short-term and long-term memory, which are used to guide pipeline generation and provide feedback for improving pipeline quality.

\leadingtext{Feedback Mechanism.} Providing feedback on pipeline performance is crucial for enhancing its quality, as generated pipelines might be incorrect. We develop several effective techniques to ensure high-quality feedback. Firstly, we utilize the pipeline execution capability to provide feedback; if a pipeline fails to execute, we can issue negative feedback. Secondly, we employ a GAN-based technique that converts the pipeline into a textual description, evaluates this description against the task description, and uses the degree of matching to provide feedback. Thirdly, we search for relevant historical tasks/pipelines and assess the similarity between this historical information and the current task and pipeline. If any inconsistencies are found, we provide negative feedback. 


\leadingtext{Agentic Pipeline Optimization.} Given a semantic pipeline with relational operators (or sub-pipeline as SQL) and semantic operators, the data analytics agent optimizes the pipeline~\cite{li2026deepeye}. This includes pushing down low-cost operators and reordering semantic operators for enhanced efficiency. These optimization techniques are highly dependent on the semantic cardinality and cost estimation of these semantic operators. To achieve this, we use the semantic catalog to assist in semantic estimation. Specifically, for a given semantic operator, we identify relevant catalog labels, fetch some samples under these labels, use the samples to estimate cardinality/cost, and then calculate the overall cardinality/cost.

\leadingtext{Pipeline Execution.} Give an optimized pipeline, we develop efficient methods for its execution. At the operator level, we design three effective techniques. First, we implement batching techniques to group multiple data items together, allowing us to invoke the LLMs more efficiently and reduce costs. Second, we utilize embedding-based techniques to filter out data items with low similarity to the query. Third, we use cascading techniques with LLMs to enhance performance by employing smaller LLMs for pruning to bypass larger LLMs. At the pipeline level, we use a vectorized execution model to improve the KV cache hit rate, thereby reducing the cost of using LLMs. We  also group the data points based on their shared prefixes in order to enhance the KV caching capabilities. 

\leadingtext{Visualization Agent.} It  transforms structured query results into clear, informative, and contextually appropriate visual representations~\cite{DBLP:journals/vldb/QinLTL20, DBLP:conf/icde/LuoQ0018, DBLP:journals/pacmmod/LuoZ00CS23}. 
It autonomously selects the most suitable chart type -- such as a bar chart for categorical comparisons, a line chart for temporal trends, or a pie chart for part-to-whole relationships. It intelligently designs the visual encoding: choosing appropriate coordinate systems, mapping data fields to visual channels ($x/y$ position, color, size), and generating meaningful legends, labels, and titles that enhance interpretability~\cite{luo2025nvbench}. The agent also considers best practices in perceptual effectiveness and avoids misleading visualizations~\cite{DBLP:journals/corr/abs-2510-22373,DBLP:journals/tvcg/LuoTLTCQ22}. By combining data semantics, task context, and visualization theory, it ensures that each visualization communicates insights accurately and intuitively to the user.

\subsection{Data Science Agent}
\label{sec:datasci}

Unlike existing methods that manually address isolated data science steps such as data discovery~\cite{DBLP:journals/vldb/QinLTL20}, data cleaning~\cite{DBLP:journals/pvldb/YangLCFCT25}, data integration, and model selection~\cite{hassan2023chatgpt,autokaggle,dsagent},  \emph{data-science agent} is capable of autonomously coordinating the entire workflow. Specifically, given an online data science task described in natural language, this stage autonomously decomposes the task into a pipeline composed of sub-tasks that align with data agent capabilities. It then selects the appropriate agent for each sub-task and dynamically refines the pipeline to ensure both efficiency and robustness.

\leadingtext{Foundational Agents for Data Science.} We develop an agent for each specific functionality in data science, including a data discovery agent, data standardization agent, data cleaning agent, model selection agent, parameter selection agent, model training agent, and model evaluation agent. We outline the functionality and description of each agent, which will be utilized in multi-agent pipeline orchestration. We  connect  common data-science tools with external ecosystems: machine learning libraries, feature engineering packages, PyData toolkits, databases, visualization platforms, and tuning frameworks. By integrating these various tools into a unified process, the data science agent enables an interpretable, adaptive, and continually evolving data science workflow that balances automation without human intervention.
Notice that our data science agent is designed to be extensible, allowing integration of new data agents through agent specification. Once integrated, the new agent can seamlessly collaborate with existing agents within our multi-agent pipeline orchestration framework.

\leadingtext{Multi-Agent Pipeline Orchestration.}  Given an online task, we start by using LLMs to decompose the task into a pipeline of multiple sub-tasks based on the foundational agent profiles. Each sub-task is assigned to an appropriate data agent~\cite{ruan2026aorchestraautomatingsubagentcreation}. We refine the pipeline iteratively with additional adjustments, such as merging or further decomposing sub-tasks. The pipeline is executed in a parallel bottom-up manner according to its topological order. To ensure robustness, we dynamically refine the pipeline based on intermediate results, including: (i) modifying sub-tasks at the agent level, and (ii) global-level re-planning, where complete intermediate results are stored in our data catalog to avoid redundant computation. Once all sub-tasks are executed, the data agent delivers the final result.

\leadingtext{Feedback Mechanism.} If a pipeline encounters errors, we provide feedback to help LLMs correct those mistakes~\cite{DBLP:journals/corr/abs-2510-17586}. To achieve this, we design several feedback mechanisms to provide effective feedback. First, we autonomously generate a data sample and use LLMs to directly generate an answer for it. This answer serves as the ground truth for evaluating the pipeline (or a sub-agent). If the pipeline's answer significantly differs from the LLM-generated answer, feedback is provided to the agent with detailed descriptions of which data is missing and which data should not be included in the answer. This strategy can also serve as feedback for the foundational agents mentioned above. For example, in the case of a data standardization agent, we can use LLMs to standardize a sample and then use the results from this sample to generate feedback. Second, we can inject some samples that include ground-truth data. If the pipeline fails to process these samples correctly, we can provide feedback with detailed descriptions. Third, we can use users' ratings, such as upvotes or downvotes, as feedback, which can be used to enhance the pipeline quality.


\begin{table}[!t]\vspace{-2em}
\centering
\setlength{\tabcolsep}{6pt}
\renewcommand{\arraystretch}{1.1}
\caption{DABStep Leaderboard Results.}\vspace{-1em}
\label{tab:dabstep}
\begin{tabular}{lcc}
\toprule
\textbf{Agent} & \textbf{Easy Acc.} & \textbf{Hard Acc.} \\
\midrule
\textbf{\oursys} & \textbf{94.44} & \textbf{50.79} \\
DS-PlaVer & 87.50 & 45.24 \\
Amity DA Agent v0.1 & 80.56 & 41.01 \\
Mphasis-IZI-Agents & 80.56 & 28.04 \\
Claude 4 Sonnet ReACT Baseline & 81.94 & 19.84 \\
Open Data Scientist & 84.72 & 16.40 \\
o4-mini Reasoning Prompt Baseline & 76.39 & 14.55 \\
Claude 3.7 Sonnet ReACT Baseline & 75.00 & 13.76 \\
o3-mini Reasoning Prompt Baseline & 72.22 & 13.76 \\
Gemini 2.5 Pro Reasoning Prompt & 66.67 & 12.70 \\
GPT 4.1 ReACT Baseline & 80.56 & 12.43 \\
o1 Reasoning Prompt Baseline & 69.44 & 11.11 \\
\bottomrule
\end{tabular}\vspace{-1em}
\end{table}

\section{Experimental Evaluation}
\label{sec:exp}

This section comprehensively evaluates our system \oursys on two challenging benchmarks that jointly cover a broad spectrum of data-centric tasks: $(i)$ DABStep, a multi-step data-analysis agent benchmark with 450 questions spanning easy and hard settings, and $(ii)$ Spider~2.0-Lite, a text-to-SQL benchmark with 547 examples designed for realistic SQL query generation across \emph{BigQuery}, \emph{Snowflake}, and \emph{SQLite} databases. Notably, DABStep evaluates multi-step reasoning over heterogeneous inputs, including unstructured documents, lookup tables, and schema-related metadata, with its hard split further stressing semantic grounding and cross-document reasoning under high-noise, multi-source settings. Spider~2.0-Lite evaluates SQL generation under schema-level noise and realistic database complexity across multiple database engines, thereby testing the agent's ability to generalize across typical SQL dialects and noisy schema environments.

For \textbf{DABStep}, we follow the official evaluation on the validated split and adopt each competitor’s public configuration as listed on the leaderboard. \oursys uses Qwen~3 as the backbone LLM and the same input documents/questions as all baselines. For fairness, we keep default decoding and tool settings unless prescribed by the leaderboard entries; no task-specific fine-tuning is applied. Accuracy is reported for both \emph{Easy} and \emph{Hard} levels, with emphasis on the latter. For \textbf{Spider 2.0-Lite}, we follow the leaderboard protocol, where all methods are ranked by execution accuracy. We compare against a broad set of representative baselines, including (1) specialized systems developed by well-known research organizations, such as Snowflake ReForce, DeepSeek LinkAlign, RSL-SQL+DeepSeek, and Claude-based Spider-Agents from Anthropic; and (2) prompt-style and agentic methods built on leading LLM families, including Qwen~3, Gemini~2.5, Claude~3.7/4, GPT o4-mini/o3-mini/o1/4.1, and DeepSeek-V3.





\subsection{Overall Accuracy}

\hi{Results on DABStep.} Table~\ref{tab:dabstep} summarizes the results. \oursys achieves the best performance on both metrics, reaching \textbf{94.44\%} (easy) and \textbf{50.79\%} (hard). Compared to the second-place system (DS-PlaVer; Gemini~2.5-Pro), \oursys improves by \textbf{+6.94} points on the easy split (94.44 vs.\ 87.50) and \textbf{+5.55} points on the hard split (50.79 vs.\ 45.24). Notably, Claude- and OpenAI-based systems perform significantly lower on the hard tasks, underscoring the importance of robust multi-step reasoning and cross-document grounding.


\begin{table}[!t]\vspace{-2em}
\centering
\setlength{\tabcolsep}{6pt}
\renewcommand{\arraystretch}{1.1}
\caption{Spider 2.0-Lite Leaderboard Results.}\vspace{-1em}
\label{tab:spiderlite}
\begin{tabular}{lc}
\toprule
\textbf{Method} & \textbf{Score} \\
\midrule
\textbf{\oursys} & \textbf{44.50} \\
ReForce + o3 (Snowflake) & 37.84 \\
ReForce + Qwen3 (Snowflake + Qwen) & 35.60 \\
Autolink + DeepSeek-R1 & 34.92 \\
RSL-SQL + o3 (DeepSeek + HUST VLR) & 33.09 \\
LinkAlign + DeepSeek-R1 (DeepSeek) & 33.09 \\
RSL-SQL + DeepSeek-R1 (DeepSeek + HUST VLR) & 30.53 \\
ReForce + o1-preview (Snowflake) & 30.35 \\
Spider-Agent + Claude-3.7-Sonnet & 28.52 \\
Spider-Agent + Claude-4-Sonnet & 27.79 \\
\bottomrule
\end{tabular}\vspace{-1em}
\end{table}

\hi{Results on Spider 2.0-Lite.} Table~\ref{tab:spiderlite} shows the Spider 2.0-Lite leaderboard. \oursys again leads with a score of \textbf{44.5}, surpassing Snowflake’s ReForce (+o3 at 37.84) by \textbf{+6.66} points and outperforming both Qwen3-enhanced and Claude-based Spider-Agents (35.6 and 28.5). The consistent advantage of \oursys across both benchmarks demonstrates its strong ability to generalize from multi-step document reasoning (DABStep) to realistic text-to-SQL tasks (Spider 2.0-Lite), reinforcing its role as a data agent framework.

\subsection{Ablation Study}

To further analyze the contribution of each core component, we conduct an ablation study on both DABStep and Spider~2.0-Lite. We categorize queries based on characteristics such as reasoning depth and data source usage complexity. 
For \textit{w/o Semantic Organization}, we remove structured schema profiling and unstructured retrieval augmentation, exposing the planner only to raw source interfaces without semantic catalogs or indexed background knowledge.
For \textit{w/o Feedback Loop}, we disable semantic validators and intermediate verification modules, retaining only runtime exception signals from downstream execution engines.
For \textit{w/o Memory Management}, we disable memory operations and only retain the context with incremental actions and feedback.

\begin{figure}[!t] \vspace{-2em}
    \centering
    \includegraphics[width=1\linewidth]{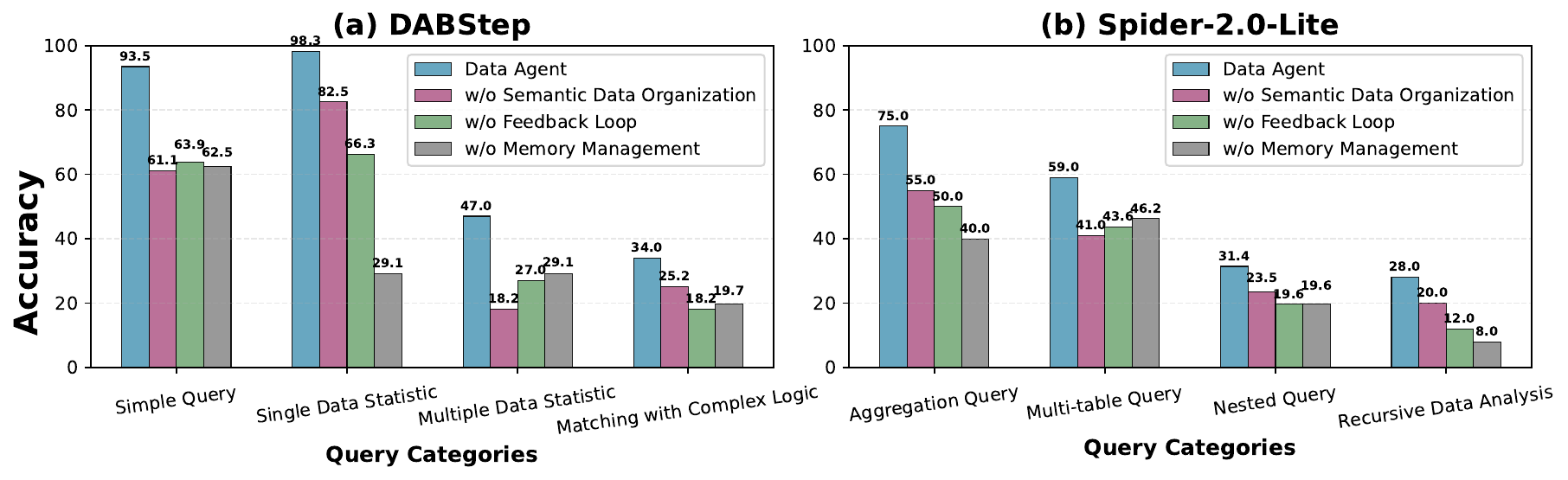}
    \vspace{-2em}
    \caption{Ablation Study Across Query Categories.}
    \label{fig:ablation}
    \vspace{-1.5em}
\end{figure}

Figure~\ref{fig:ablation} shows that all three components consistently contribute to performance across both DABStep and Spider~2.0-Lite, while their impact varies by query characteristics. 

\hi{Without Semantic Organization.} Removing Semantic Organization leads to a performance drop, especially on categories that require cross-source semantic alignment or retrieval over large, noisy data spaces, such as \textit{multiple data statistics} in DABStep and \textit{multi-table queries} in Spider~2.0-Lite. This confirms that semantic catalogs and indexes are critical for mapping high-level user intents to relevant data in large and noisy data sources.

\hi{Without Feedback Loop Management.} Disabling Feedback Loop causes more noticeable degradation on queries with deeper reasoning chains (e.g., {complex logic matching}, {recursive data analysis}). This suggests the feedback mechanisms are particularly effective at detecting latent logical deviations that are increasingly difficult for LLMs to find during planning as query complexity and reasoning depth grow.

\hi{Without Memory Management.} The effect of removing Memory Management is more noticeable on all but the simplest queries. As query complexity increases, multi-turn pipeline orchestration introduces more intermediate decisions and feedback signals, making the context lengthy and diluting the LLM's attention. The memory manager preserves the most relevant intermediate decisions and feedback signals during multi-turn pipeline orchestration, and also enables the reuse of dataset-specific knowledge and previously successful pipeline patterns, thereby keeping the planning context compact and allowing the agent to solve complex tasks more efficiently.

Additionally, the proactive capabilities of \oursys are partially validated through the ablation study: the \textit{w/o Memory Management} variant, which disables the evolving memory and pattern reuse, shows consistent performance degradation across both benchmarks, confirming the memory-driven proactive adaptation contributes meaningfully to system effectiveness. More advanced proactive capabilities, such as fully autonomous problem identification and self-initiated corrective workflows without human prompts, remain open challenges. 

Overall, the ablation results align well with the intended responsibilities of the three modules, showing that the gains of \oursys do not come from a single component but from their complementary interaction.

\begin{figure}[!t]\vspace{0em}
    \centering
    \includegraphics[width=1\linewidth]{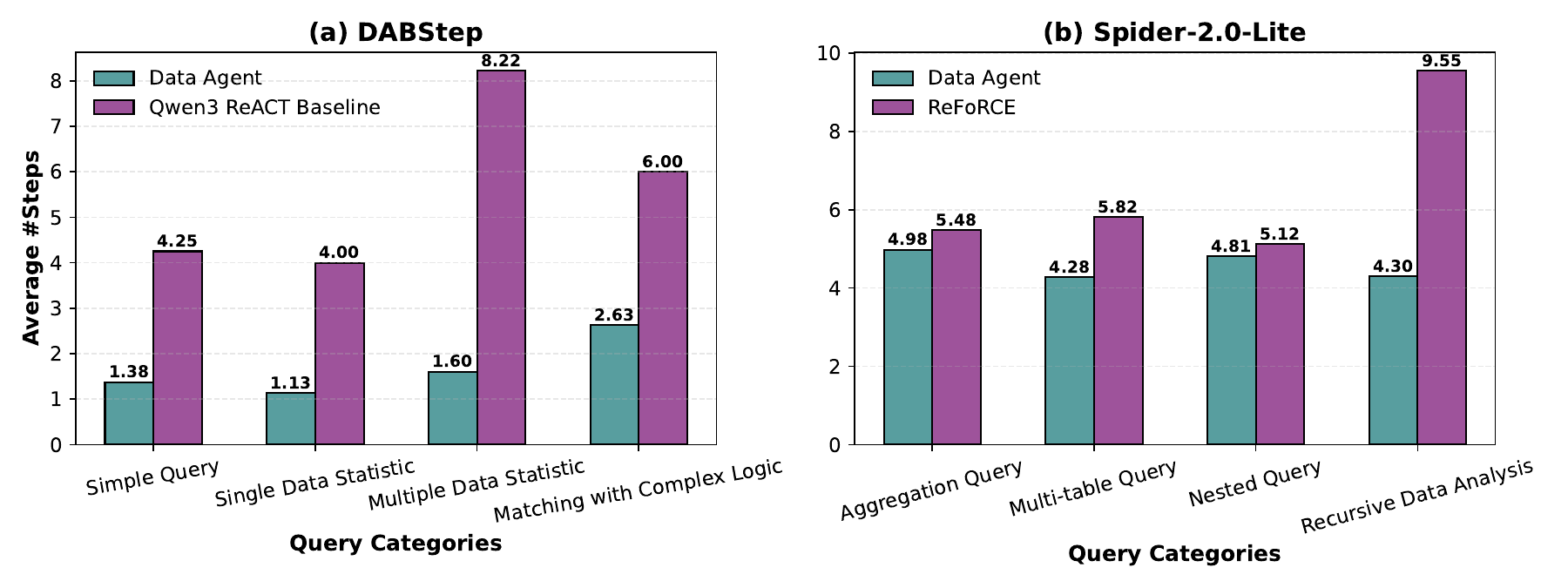}\vspace{-1em}\vspace{-.5em}
    \caption{Comparison of Iteration Steps.}
    \label{fig:steps_compare}\vspace{-1.5em}
\end{figure}

\vspace{-.25em}
\subsection{Cost and Efficiency Analysis}
\vspace{-.25em}

We analyze the token and efficiency benefits brought by \oursys from end-to-end latency and token consumption.

\hi{Execution Time Reduction.} Table~\ref{tab:time_compare} reports the average execution time per query on DABStep and Spider~2.0-Lite. \oursys consistently reduces the total execution time compared with the corresponding baselines, achieving a $2.32\times$ speedup on DABStep and a $1.45\times$ speedup on Spider~2.0-Lite. The latency reduction mainly comes from the complementary effects of the three core components.
First, the Semantic Data Organization component improves data understanding and source grounding, allowing the planner to identify relevant tables, documents, and attributes earlier and thus reducing unnecessary trial-and-error exploration.
Second, the Feedback Loop detects semantic deviations and logical inconsistencies during orchestration, preventing erroneous plans from propagating into expensive downstream execution failures.
Third, the Memory Management component accumulates reusable dataset-specific procedures, successful workflow patterns, and historical corrections, enabling the system to reuse previously effective orchestration strategies instead of repeatedly exploring similar search spaces from scratch.

\begin{table}[!t]\vspace{-3em}
\centering
\caption{End-to-end execution time comparison.}
\label{tab:time_compare}
\vspace{-1em}
\setlength{\tabcolsep}{2pt} 
\begin{tabular}{lccc}
\toprule
Dataset & Data Agent & Baseline & Speedup \\
\midrule
DABStep & 2.8 min & 6.5 min (Qwen3 ReACT Baseline) & 2.32$\times$ \\
Spider~2.0-Lite & 5.1 min & 7.4 min (ReFoRCE) & 1.45$\times$ \\
\bottomrule
\end{tabular}\vspace{-1em}
\vspace{-.5em}
\end{table}

\hi{Iteration Step Reduction.} 
To better understand where the efficiency gain comes from, Figure~\ref{fig:steps_compare} further compares the average number of reasoning-and-execution steps across query categories.
Here, a \textit{step} denotes one complete orchestration or execution iteration in the orchestration pipeline, including intermediate verification and potential local repair. This metric captures the overall interaction complexity of the full data analytics workflow. We have two key observations. First, \oursys consistently takes fewer steps across nearly all query categories, showing that the architecture reduces unnecessary exploration and repair cycles throughout the full pipeline. Second, as query difficulty increases, the step growth of \oursys is flatter than the baselines, especially on categories involving multiple data sources, nested logic, or complex analytical goals. This indicates that the combined effects of semantic data understanding, early feedback-based correction, and reusable memory effectively control the expansion of orchestration horizons in complex scenarios.

\begin{table}[!t]
\centering
\caption{Token Usage Comparison on DABStep (ReACT is Qwen3 ReACT).}
\label{tab:token_dabstep}
\vspace{-1em}
\setlength{\tabcolsep}{3.8pt} 
\scriptsize
\renewcommand{\arraystretch}{1.2}
\begin{tabular}{l|cc|cc}
\toprule
\multirow{2}{*}{Category} 
& \multicolumn{2}{c|}{Input Tokens (k)} 
& \multicolumn{2}{c}{Output Tokens (k)} \\
\cmidrule(r){2-3} \cmidrule(r){4-5}
& \oursys & ReACT & \oursys & ReACT \\
\midrule
Simple Query            & 16.9 & 33.3 & 0.91 & 2.93 \\
Single Data Statistic       & 17.0 & 120.3 & 0.85 & 2.56 \\
Multiple data Statistic        & 18.0 & 110.2 & 1.10 & 3.67 \\
Matching with Complex Logic & 27.6 & 178.5 & 2.94 & 13.68 \\
\bottomrule
\end{tabular}
\vspace{-1em}
\end{table}

\begin{table}[!t]\vspace{-1em}
\centering
\caption{Token Usage Comparison on Spider~2.0-Lite.}\vspace{-1em}
\label{tab:token_spider}
\setlength{\tabcolsep}{3.8pt} 
\scriptsize
\renewcommand{\arraystretch}{1.2}
\begin{tabular}{l|cc|cc}
\toprule
\multirow{2}{*}{Category} 
& \multicolumn{2}{c|}{Input Tokens (k)} 
& \multicolumn{2}{c}{Output Tokens (k)} \\
\cmidrule(r){2-3} \cmidrule(r){4-5}
& \oursys & ReFoRCE & \oursys & ReFoRCE \\
\midrule
Aggregation Query                    & 10.8 & 71.2 & 2.98 & 11.32 \\
Multi-table Query                    & 18.5 & 88.4 & 4.79 & 6.24 \\
Nested Query                         & 20.4 & 64.7 & 4.82 & 8.13 \\
Recursive Data Analysis        & 18.5 & 52.8 & 6.76 & 13.21 \\
\bottomrule
\end{tabular}
\vspace{-1.5em}
\end{table}

\hi{LLM Cost Reduction.} Tables~\ref{tab:token_dabstep} and~\ref{tab:token_spider} further compare token usage. \oursys uses fewer tokens, and the token growth remains more stable as task difficulty increases. On DABStep, \oursys reduces input tokens from 33.3k (by Qwen3 ReACT Baseline) to 16.9k on simple queries, and from 178.5k to 27.6k on matching tasks with complex logic, achieving up to a 6.5$\times$ reduction. This difference is mainly because the baseline approach tends to operate over larger raw contexts and rely on repeated exploration during multi-step reasoning, leading to substantial token consumption. In contrast, \oursys leverages semantic data organization and memory to focus on relevant data and reuse prior knowledge, resulting in more compact and efficient context usage.

\section{Related Work and Open Problems}
\label{sec:stages}


We classify the capabilities of data agents into five levels (L1–L5) according to their degree of autonomy and learning ability. \textbf{L1} designates the data agent as a human-assisted copilot, with humans taking the lead while the agent assists users through prompt-based tool invocation and RAG~\cite{DBLP:conf/icde/ZhangLS25,stage0rewrite,stage0htap,stage0clean,stage0entitymatching,stage0schemamatching,stage0tablellm,stage0schemadiscovery}. \textbf{L2} facilitates task-oriented agent automation, where users specify high-level intents, and the agent independently manages task decomposition, planning, orchestration, reasoning, and decision-making~\cite{stage2lambda,stage2infiagent,stage2nvagent,stage2alphasql,stage2tablemaster,stage2tablecritic,stage2pipelines,stage2dbot,tang2026straptor,DBLP:journals/pacmmod/ZhangFFYLLWZ26,DBLP:conf/sigmod/LuoLFT26,DBLP:journals/pvldb/LiSPWXLN25}.  \textbf{L3} makes a shift to an agent-led workflow and positions the data agent as the lead, continuously enhancing its capabilities through iterative human feedback, continuous learning, and collaboration with multiple agents. This allows the system to adapt, generalize across tasks, and improve with experience~\cite{stage3gaussmaster,stage3unify,stage3multimodal,stage3ds,stage3chat2data,stage3agenticdata,stage3idatalake,stage3palimpzest,aop,DBLP:journals/pvldb/SunZLYFZ25,CIDRDS,tang2025llmagentasdataanalystsurvey,sigmod26semanticcard,DBLP:journals/pvldb/HuLLWL26}.  \textbf{L4} proactively initiates tasks, autonomously detects events, and adaptively addresses the events with complex collaboration and short-term and long-term memory management techniques. \textbf{L5} represents a fully autonomous agent capable of operating independently, performing tasks from start to finish without human intervention.



Although this paper proposes a data agent system and develops key techniques to enhance result quality and reduce LLM token costs, several open challenges remain.






\leadingtext{Self-Evolving Data Agents.}  Self-evolving data agents aim to improve their ability to discover, query, analyze, and manage data through accumulated experience. However, achieving reliable and sustained improvement raises several open challenges. First, reliable feedback and credit assignment remain difficult. A successful execution does not necessarily imply a correct analysis: an agent may generate executable SQL while using inappropriate joins, misleading aggregations, or unsupported assumptions. Feedback must distinguish execution success from semantic correctness and identify which decisions contributed to an outcome. Second, safe and stable evolution requires balancing adaptation with preservation of established capabilities. Updates to prompts, memories, tools, or policies may improve one workload while degrading another. Determining when to update, how to validate changes, and when to roll back remains challenging, particularly under changing schemas and data distributions. Third, memory quality and transferability are unresolved. Agents must determine which experiences to retain, reconcile contradictory observations, and avoid treating context-specific solutions as general rules. Effective reuse requires tracking provenance, applicability conditions, and expiration, while respecting privacy and access boundaries.

\leadingtext{Advanced Proactive Capabilities.} \oursys supports basic proactive capabilities, including adapting semantic data catalogs to new data sources, adjusting vector indexes to shifting query distributions, and leveraging long-term memory to reuse effective pipeline patterns. However, more advanced proactive behaviors—such as autonomously detecting events, launching tasks without user prompts, and executing fully self-directed workflows—remain significant research hurdles. Addressing these challenges will require breakthroughs in continuous monitoring, autonomous goal formulation, and robust self-healing mechanisms to enable our data agents to operate independently and adaptively in dynamic environments.%

\leadingtext{Data Security and Privacy Issue.} Deploying data agents in industrial settings poses some challenges. (1) Data Security and Privacy: Robust measures like data masking, local model deployment, and strict access control are needed to protect sensitive information such as personally identifiable information (PII) and prevent data leakage during LLM calls or memory sharing. (2) Regulatory Compliance: Industries must adhere to regulations like GDPR, requiring systems to ensure auditability and traceability. (3) Domain Knowledge Integration: Many tasks depend on specialized domain knowledge not captured during LLM training, making efficient integration and updating of this knowledge critical. (4) Robustness under Extreme Conditions: Adversarial inputs and distribution shifts demand self-healing mechanisms and enhanced resilience.


\leadingtext{Multimodal Data Agents.} Analysts regularly work across diverse modalities, including textual reports, structured tables, time-series logs, visual charts, source code, videos, audios, and knowledge graphs. A key challenge is the semantic and structural misalignment between these modalities.

\vspace{-.5em}
\section{Conclusion}
\label{sec:conclusion}

We propose a data agent paradigm to revolutionize data systems by autonomously and proactively handling data-related tasks with semantic capabilities. Our data agent system is designed with components for semantic data organization, semantic operators, agentic pipeline orchestration, feedback mechanisms, memory management, and proactive adaptation. We develop a data analytics agent and a data science agent. Experimental results demonstrate that our data agents significantly outperform state-of-the-art methods. 




\section*{Acknowledgment}
This paper was supported by NSF of China (62525202, 62232009), National Key R\&D Program of China (2023YFB4503600), and Fundamental and Interdisciplinary Disciplines Breakthrough Plan of the Ministry of Education of China(JYB2025XDXM509).

\vspace{-.5em}
\scriptsize
\bibliographystyle{abbrv}
\bibliography{ref/DA}

\end{CJK*}
\end{document}